\documentclass[a4paper,11pt]{article}

\usepackage[english]{babel}
\usepackage{setspace}

\usepackage[a4paper,tmargin=3.5truecm,bmargin=4truecm,rmargin=2.5truecm,
lmargin=3.2truecm,twoside,verbose=true]{geometry}
\usepackage{cancel,graphicx}
\usepackage{hyperref}
\usepackage{amsmath,amssymb,slashed}

\usepackage{amsmath,amssymb}
\usepackage[amsmath, hyperref, thmmarks]{ntheorem}
\numberwithin{equation}{section}

\allowdisplaybreaks[1]
\newcommand\bbone{{\mathbb{I}}}

\renewenvironment{thebibliography}[1]
         {\section*{References}\frenchspacing\small
          \begin{list}{[\arabic{enumi}]}
         {\usecounter{enumi}\parsep=2pt\topsep 0pt
         \settowidth{\labelwidth}{[#1]}
         \leftmargin=\labelwidth\advance\leftmargin\labelsep
         \rightmargin=0pt\itemsep=1pt\sloppy}}{\end{list}}

\theoremsymbol{}
\theorembodyfont{\slshape}
\theoremheaderfont{\normalfont\bfseries}
\theoremseparator{}

\theorembodyfont{\upshape}
\theoremsymbol{\ensuremath{\blacklozenge}}

\theoremstyle{nonumberplain}
\theoremheaderfont{\scshape}
\theorembodyfont{\normalfont}
\theoremsymbol{\ensuremath{\blacksquare}}

\qedsymbol{\ensuremath{_\blacksquare}}
\theoremclass{LaTeX}

\title{Twisted BRST symmetry in gauge theories on $\kappa$-Minkowski }
\author{Philippe Mathieu$^a$, Jean-Christophe Wallet$^b$}
\begin{document}

\date{}
\maketitle
\vspace*{-1cm}

\begin{center}
\textit{$^a$Department of Mathematics, 
University of Notre Dame, Notre Dame, IN 46556, USA.} \\
\textit{$^b$IJCLab, Universit\'e Paris-Saclay, CNRS/IN2P3, 91405 Orsay, France.}  \\
\textit{}\\
\bigskip
 e-mail:
\texttt{pmathieu@nd.edu, jean-christophe.wallet@th.u-psud.fr}\\[1ex]

\end{center}

\begin{abstract} 
 Algebraic properties of the BRST symmetry associated to the twisted gauge symmetry occurring in the $\kappa$-Poincar\'e invariant gauge theories on the $\kappa$-Minkowski space are investigated. We find that the BRST operation associated to the gauge invariance of the action functional can be continuously deformed together with its corresponding Leibniz rule, into a nilpotent twisted BRST operation, leading to a twisted BRST symmetry algebra which may be viewed as a noncommutative analog of the usual Yang-Mills BRST algebra.
\end{abstract}

\newpage


\section{Introduction.}

Some attention has been paid to Noncommutative Field Theories (NCFT) on $\kappa$-Minkowski spaces for more than two decades, see e.g. \cite{habsb-impbis}-\cite{old-gauge} and references therein. The $\kappa$-Minkowski space $\mathcal{M}_\kappa^d$ can be described in its simplest version as the enveloping algebra of the solvable Lie algebra of  ``noncommutative coordinates{\footnote{Here, $x_0,\ x_i$ are self-adjoint operators; it is often assumed that  $\kappa\sim\mathcal{O}(M_{Planck})$ in $4$-dimensional.}}'' defined by $[x_0,x_i]=\frac{i}{\kappa}x_i,\ \ [x_i,x_j]=0,\ \ i,j=1,\cdots, (d-1)$. This attention has been increased by the belief that $\kappa$-Minkowski spacetimes may capture salient features of the quantum spacetime underlying Quantum Gravity at least in some limit. This is partly motivated by observing \cite{maatz} that the $\left(2+1\right)$-dimensional quantum gravity coupled to matter, upon integrating out the gravitational degrees of freedom, yields a field theory invariant under a deformation of the Poincar\'e algebra, the $\kappa$-Poincar\'e algebra $\mathcal{P}_\kappa^d$ \cite{luk1}. But this latter quantum algebra is the building block of the $\kappa$-Minkowski space $\mathcal{M}_\kappa^d$ which is nothing but the dual of a subalgebra of $\mathcal{P}_\kappa^d$, the ``algebra of deformed translations'' $\mathcal{T}_\kappa^d$, as first evidenced in \cite{majid-ruegg} from the Hopf algebra bicrossproduct structure of $\mathcal{P}_\kappa^d$ with covariant coaction on $\mathcal{M}_\kappa^d$. For a general review, see e.g. \cite{luk2}. For useful properties of $\mathcal{P}_\kappa^d$, see Appendix \ref{apendixB}.\\

Extrapolating the validity of the above observation to $\left(3+1\right)$-dimensional has somehow reinforced the idea that $\kappa$-Minkowski and $\kappa$-Poincar\'e structures are of some relevance to understand the behavior of the $\left(3+1\right)$-dimensional quantum gravity in some regime near the Planck scale, triggering a huge number of works on NCFT on $\kappa$-Minkowski spaces. Comparatively to the noncommutative gauge theories on Moyal spaces $\mathbb{R}^{2n}_\theta$ or on $\mathbb{R}^3_\lambda$ \cite{biko1} for which classical and/or quantum properties have been examined to some extent \cite{reviewgaugemoyal}-\cite{wal-16}, it appears that noncommutative gauge theories on $\kappa$-Minkowski spaces have been much less investigated although past works \cite{old-gauge} have opened the way for their exploration.\\

Recently, we have taken advantage of the convenient star-product used in \cite{PW2018} to characterize the classical properties of $\kappa$-Poincar\'e invariant gauge theories on $\kappa$-Minkowski spaces \cite{MW20}. It is given{\footnote{Spacelike (resp. timelike) coordinates refer to Latin $i,j,\hdots=1,2,...,(d-1)$  (resp. $0$)  indices. $x:=(x_\mu)=(x_0,\vec{x})$ and $x.y:=x_\mu y^\mu=x_0y_0+\vec{x}\vec{y}$. The Fourier transform of $f\in L^1(\mathbb{R}^d)$ is $(\mathcal{F}f)(p):=\int d^dx\ e^{-i(p_0x_0+\vec{p}.\vec{x})}f(x)$ with inverse $\mathcal{F}^{-1}$, $\bar{f}$ its complex conjugate. $\mathcal{S}_c$ is the space of Schwartz functions with compact support in the first variable.}}, together with the involution, by
\begin{align}
(f\star g)(x)&=\int \frac{dp^0}{2\pi}dy_0\ e^{-iy_0p^0}f(x_0+y_0,\vec{x})g(x_0,e^{-p^0/\kappa}\vec{x})  \label{starpro-4d},\\
f^\dag(x)&= \int \frac{dp^0}{2\pi}dy_0\ e^{-iy_0p^0}{\bar{f}}(x_0+y_0,e^{-p^0/\kappa}\vec{x})\label{invol-4d},
\end{align}
for any $f,g$ in a suitable multiplier algebra{\footnote{For a full characterization of this algebra multiplier, see \cite{DS}. It is the algebra of smooth functions with polynomial bounds together with all their derivatives and such that their inverse Fourier transform w.r.t. the $x_0$ variable is compactly supported.}} which we will simply denote by $\mathcal{M}_\kappa^d$. Recall that this product stems from a mere combination of the Weyl-Wigner quantization map with the convolution algebra of the affine group $\mathbb{R}\ltimes\mathbb{R}^{(d-1)}$. The $\kappa$-Poincar\'e invariance of the functional action forces the trace in the action to be the usual Lebesgue integral which however is no longer cyclic with respect to the above star-product. As a result, a twist shows up upon cyclic permutation of the factors inside the trace. The twist, called modular twist \cite{PW2018}, and depending on the dimension $d$ that appears in the affine group from which $\mathcal{M}_\kappa^d$ is built, prevents the factors stemming from the gauge transformation to compensate each other. In \cite{MW20}, we have shown that the modular twist effect can be entirely neutralized, leading to a $\kappa$-Poincar\'e invariant and gauge invariant functional action with physically acceptable commutative limit, provided the $\kappa$-Minkowski space is $5$-dimensional{\footnote{Starting from another star-product, a $4$-dimensional gauge theory on $\kappa$-Minkowski has been obtained in \cite{russovit}, which however does not give rise to a standard commutative limit. }}. This can be achieved thanks to the existence of a unique twisted noncommutative differential calculus based on a family of twisted derivations of  $\mathcal{T}_\kappa^d$ \cite{MW20-bis}. Some physical properties of the $4$-dimensional effective theory obtained from standard compactification scenarii have been analyzed and confronted to recent data from collider experiments and Gamma Ray Burst photons in \cite{MW20-bis}.\\

In this paper, we study the algebraic characterization of the BRST symmetry associated to the twisted gauge symmetry ruling the above $\kappa$-Poincar\'e invariant gauge theories. The BRST symmetry for noncommutative gauge theories was considered for the first time in the literature in \cite{ga} for the Moyal space and in \cite{bu} for the noncommutative torus. While the algebraic structure of the BRST symmetry for gauge theories on Moyal spaces, $\mathbb{R}^3_\lambda$ or on the noncommutative torus follows closely, at least formally, the one of (commutative) Yang-Mills theory, the appearance of a twisted gauge symmetry combined with a twisted differential calculus, both mandatory to insure the gauge invariance \cite{MW20}, modifies the algebraic structure coding the BRST symmetry associated to the gauge theories considered here. We find in particular that the BRST operation related to the gauge invariance of the action functional can be continuously distorted into a twisted BRST operation, preserving nilpotency and anticommutativity with $\bf{d}$, distorting continuously the Leibniz rule. This twisted BRST operation gives rise to a twisted BRST symmetry algebra, resulting in two nilpotent operations related to the gauge symmetry.\\

To make the paper self-contained, the useful properties of the twisted differential calculus and of the twisted connection are collected in subsections \ref{section21} and \ref{section22}. The BRST operation leaving the action functional invariant is presented in subsection \ref{section23}. In subsection \ref{section31}, we recall the algebraic structure of the BRST algebra for (commutative) Yang-Mills theory while the basics of the Weil algebra are collected in the appendix \ref{weil}. The twisted BRST symmetry algebra is considered in subsection \ref{section32}. In section \ref{section4}, we conclude.

\section{BRST symmetry for gauge theories on $\kappa$-Minkowski.}

\label{section2}

\subsection{Twisted differential calculus.}

\label{section21}

The relevant family of twisted derivations of $\mathcal{T}_\kappa^d$ is given by
\begin{equation}
X_0=\kappa\mathcal{E}^\gamma(1-\mathcal{E})\,\mbox{ and }\,X_i=\mathcal{E}^\gamma P_i\,\mbox{ for }\, i=1,...,(d-1)\label{tausig-famil}
\end{equation}
where for the moment we do not fix $d$ to its special value $d=5$ \cite{MW20} and $\gamma$ is a real parameter. One easily verifies that 
\begin{equation}
[X_\mu,X_\nu]:=X_\mu X_\nu-X_\nu X_\mu=0\label{abel-deriv}
\end{equation}
so that the $X_\mu$'s form an Abelian Lie algebra for the usual commutator $\left[\cdot,\cdot\right]$, denoted below as $\mathfrak{D}_\gamma$.\\
Recall that the $X_\mu$ belong to a particular type of twisted derivations sometimes known in the mathematical literature as ($\tau$,$\sigma)$-derivations where the morphisms $\tau$ and $\sigma$ twist the standard Leibniz as we will show below. For recent applications of these twisted derivations in Ore extensions and Hom-Lie algebras, see e.g. \cite{ore,Hom-Lie}.\\
In the case of \eqref{tausig-famil}, the twisted Leibniz rule is given by
\begin{equation}
X_\mu(a\star b)=X_\mu(a)\star (\mathcal{E}^\gamma\triangleright b)+ (\mathcal{E}^{1+\gamma}\triangleright a)\star X_\mu(b),\label{tausigleibnitz}
\end{equation}
for any $a,b\in\mathcal{M}_\kappa^d$ (hence $\tau=\mathcal{E}^{\gamma}$ and $\sigma=\mathcal{E}^{1+\gamma}$). Eqn. \eqref{tausigleibnitz} stems from the definition of the $X_\mu$ \eqref{tausig-famil} combined with the structure of the coproduct equipping $\mathcal{P}_\kappa^d$. Furthermore, the algebra $\mathfrak{D}_\gamma$ verifies  
\begin{equation}
(X.z)(a):=X(a)\star z=z\star X(a)=(z.X)(a)
\end{equation}
for any $a\in\mathcal{M}_\kappa^d$ and any $z\in Z(\mathcal{M }_\kappa^d)$, the center of $\mathcal{M }_\kappa^d$, thus exhibiting a structure of bimodule over $Z(\mathcal{M }_\kappa^d)$.\\

We first introduce the twisted differential calculus{\footnote{For untiwsted noncommutative differential calculus, see \cite{mdv} and references therein. }} based on the algebra $\mathfrak{D}_\gamma$ of twisted derivations \eqref{tausig-famil} which underlies the whole framework ruling the $\kappa$-Poincar\'e invariant gauge theories developed in \cite{MW20,MW20-bis}. \\
In this differential calculus, $n$-forms are defined from the linear space $\Omega^n(\mathfrak{D}_\gamma)$ of $n$-linear antisymmetric forms, where linearity of forms holds w.r.t. $\mathcal{Z}(\mathcal{M}_\kappa^d)$. Then, for any $n$-form $\alpha\in\Omega^n(\mathfrak{D}_\gamma)$, one has $\alpha:\mathfrak{D}_\gamma\to\mathcal{M}_\kappa^d$ with
\begin{eqnarray}
\alpha(X_1,X_2,...,X_n)&\in&\mathcal{M}_\kappa^d,\label{formule1}\\
\alpha(X_1,X_2,...,X_n.z)&=&\alpha(X_1,X_2,...,X_n)\star z\label{formule2},
\end{eqnarray}
for any $z$ in $Z(\mathcal{M }_\kappa^d)$ and any $X_1,...X_n\in\mathfrak{D}_\gamma$.\\
 We now define
\begin{equation}
\Omega^\bullet:=\bigoplus_{n=0}^{d}\Omega^n(\mathfrak{D}_\gamma), \label{omegabullet}
\end{equation}
 with $\Omega^0(\mathfrak{D}_\gamma)=\mathcal{M}_\kappa^d$. This linear space inherits a structure of associative algebra when equipped with the product of forms defined for any $\alpha\in\Omega^p(\mathfrak{D}_\gamma)$, $\beta\in\Omega^q(\mathfrak{D}_\gamma)$ by
\begin{equation}
\alpha\times\beta\in\Omega^{p+q}(\mathfrak{D}_\gamma)
\end{equation}
with
\begin{align}
\nonumber
&(\alpha\times\beta)(X_1,...,X_{p+q})\\
&\qquad\qquad=\frac{1}{p!q!}\sum_{s\in\mathfrak{S}(p+q)}(-1)^{\text{sign}(s)}\alpha(X_{s(1)},...,X_{s(p)})\star \beta(X_{s(p+1)},...,X_{s(q)})\label{ncwedge1},
\end{align}
in which $\mathfrak{S}(p+q)$ is the symmetric group of a set of $p+q$ elements, $\text{sign}(s)$ is the signature of the permutation $s$. Finally, the differential is defined by 
\begin{equation}
{\bf{d}}:\Omega^p(\mathfrak{D}_\gamma)\to\Omega^{p+1}(\mathfrak{D}_\gamma),\ \forall\,p\in\left\lbrace 0,...,(d-1) \right\rbrace
\end{equation}
with
\begin{equation}
\left({\bf{d}}\alpha\right)\left(X_1,X_2,...,X_{p+1}\right)
=\sum_{i=1}^{p+1}(-1)^{i+1}X_i\left(\alpha(X_1,...,\vee_i,...,X_{p+1})\right), \label{ncwedge2}
\end{equation}
where the symbol $\vee_i$ indicates the omission of $X_i$. The differential satisfies
\begin{equation}
{\bf{d}}^2=0. \label{nilpot}
\end{equation}
It can be easily verified that the differential ${\bf{d}}$ satisfies the following twisted Leibniz rule
\begin{equation}
{\bf{d}}(\alpha\times\beta)={\bf{d}}\alpha\times\mathcal{E}^\gamma(\beta)+(-1)^{\delta(\alpha)}\mathcal{E}^{1+\gamma}(\alpha)\times{\bf{d}}\beta\label{leibnitz-form},
\end{equation}
where $\delta({\alpha})$ is the form-degree of $\alpha$ and $\mathcal{E}^x(\alpha)$ is defined for any real number $x$ and any form $\alpha$ with degree $n$ by $\mathcal{E}^x(\alpha)\in\Omega^n(\mathfrak{D}_\gamma)$ with
\begin{equation}
\mathcal{E}^x(\alpha)(X_1,...X_n)=\mathcal{E}^x\triangleright(\alpha(X_1,...X_n)).
\end{equation}
Then, the triple $(\Omega^\bullet,\times,{\bf{d}})$ defines a graded differential algebra.\\

\noindent At this stage, three comments are in order: 
\begin{enumerate}
\item Given the algebra $\mathfrak{D}_\gamma$ related to \eqref{tausig-famil}, the definition of the elements of $\Omega^\bullet$ implies that the maximal degree of the forms is equal to $d$, stemming simply from the antisymmetry of forms.
\item One has $\alpha\times\beta\ne(-1)^{\delta(\alpha)\delta(\beta)}\beta\times\alpha$, contrary to what happens for the standard commutative (de Rham) differential calculus. In particular, given a $1$-form $A$, one has $A\times A\ne 0$ as it can be easily verified by using \eqref{ncwedge1}. It follows that the differential algebra $(\Omega^\bullet,\times,{\bf{d}})$ is {\it{not}} graded commutative.
\item We recall for further use that the $X_\mu$ are self-adjoint operators w.r.t. the Hilbert product $\langle a,b \rangle=\int d^dx\ a^\dag\star b$, i.e.
$\langle a,X_\mu(b) \rangle=\langle X_\mu(a),b \rangle$ and that $\int d^dx$ is a twisted trace w.r.t. the star product \eqref{starpro-4d}, namely \cite{PW2018}
\begin{equation}
\int d^dx\ a\star b=\int d^dx\ (\mathcal{E}^{d-1}\triangleright b)\star a\label{twistedtrace}
\end{equation}
for any $a,b\in\Omega^0(\mathfrak{D}_\gamma)$.

\end{enumerate}

\subsection{Twisted connection and curvature.}

\label{section22}

Let $\mathbb{E}$ be a right-module over $\mathcal{M}_\kappa^d$, {\it{assumed in the sequel to be one copy of $\mathcal{M}_\kappa^d$}}, i.e. 
\begin{equation}
\mathbb{E}\simeq\mathcal{M}_\kappa^d. \label{onecopy}
\end{equation}
We will nevertheless use separate symbols for the algebra and the module in the sequel when necessary. \\
In the following, the action of $\mathcal{M}_\kappa^d$ on $\mathbb{E}$ is assumed to be given by 
\begin{equation}
m\triangleleft a=m\star a.\label{aktion}
\end{equation}
The twisted connection is defined \cite{MW20} as a map 
\begin{equation}
\nabla_{X_\mu}:\mathbb{E}\to\mathbb{E},\ \ \forall\, X_\mu\in\mathfrak{D}_\gamma\label{map1}
\end{equation}
satisfying
\begin{eqnarray}
\nabla_{X_\mu+X^\prime_\mu}(m)&=&\nabla_{X_\mu}(m)+\nabla_{X^\prime_\mu}(m)\label{sigtaucon1}\\
\nabla_{z.X_\mu}(m)&=&\nabla_{X_\mu}(m)\star z\label{sigtaucon2}\\
\nabla_{X_{\mu}}(m\star a)&=&\nabla_{X_{\mu}}(m)\star(\mathcal{E}^{\gamma}\triangleright a)+(\mathcal{E}^{\gamma+1}\triangleright m)\star X_{\mu}(a),\label{sigtauconbis1}
\end{eqnarray}
for any $m\in\mathbb{E}$, $X_\mu,X^\prime_\mu\in\mathfrak{D}_\gamma$, $z\in Z(\mathcal{M }_\kappa^d)$, $a\in\mathcal{M}_\kappa^d$. Note that in  \eqref{sigtauconbis1} the factor $(\mathcal{E}^{\gamma+1}\triangleright m)$ in the second term must be understood as a morphism, say $\tilde{\beta}:\mathbb{E}\to\mathbb{E}$, whose action on the module is simply defined by $\tilde{\beta}(m)=\mathcal{E}^{\gamma+1}\triangleright m$ for any $m$ in $\mathbb{E}\simeq\mathcal{M}_\kappa$. \\

For $m=\bbone$, \eqref{sigtauconbis1} yields 
\begin{equation}
\nabla_{X_{\mu}}(a)=A_\mu\star(\mathcal{E}^{\gamma}\triangleright a)+X_{\mu}(a)\label{cov-der}
\end{equation}
where we set 
\begin{equation}
A_\mu:=\nabla_{X_{\mu}}(\bbone), \ \ \nabla_\mu:=\nabla_{X_\mu}\label{cestamiou}
\end{equation}
thus introducing the gauge potential. More generally, we will denote the evaluation of any form $\alpha $ on any derivation $X_\mu\in\mathfrak{D}_\gamma$ as $\alpha_\mu:=\alpha(X_\mu)$.\\
The map defined by \eqref{cov-der} easily extends to a map 
\begin{equation}
\nabla:\mathbb{E}\to\mathbb{E}\otimes\Omega^1(\mathfrak{D}_\gamma)\label{cestamioubis}
\end{equation}
such that
\begin{equation}
\nabla(a)=A\star a+ 1\otimes {\bf{d}}a\label{forms-alph1}
\end{equation}
with $A\in\Omega^1(\mathfrak{D}_\gamma)$ is the $1$-form gauge connection, ${\bf{d}}$ is still given by \eqref{ncwedge2} with ${\bf{d}}^2=0$.\\

The curvature can be defined from the map $\mathcal{F}(X_\mu,X_\nu):=\mathcal{F}_{\mu\nu}:\mathbb{E}\to\mathbb{E}$ such that
\begin{equation}
\mathcal{F}_{\mu\nu}=\mathcal{E}^{1-\gamma}(\nabla_\mu\mathcal{E}^{-1-\gamma}\nabla_\nu-\nabla_\nu\mathcal{E}^{-1-\gamma}\nabla_\mu)\footnote{This definition is not exactly the same as the one we used in \cite{MW20} and \cite{MW20-bis}. In those references, the curvature was a morphism of twisted module, while here we introduced a twist in order to turn the curvature into a morphism of (non-twisted) module. As a consequence, the expression of the classical actions will be slightly different also totally equivalent.}\label{decadix}.
\end{equation}
By using eqn.\eqref{sigtauconbis1}, it can be easily verified that $\mathcal{F}_{\mu\nu}$ \eqref{decadix} is a morphism of module, namely one has
\begin{equation}
\mathcal{F}_{\mu\nu}(m\star a)=\mathcal{F}_{\mu\nu}(m)\star a\label{morphimod}
\end{equation}
for any $m\in\mathbb{E}$, $a\in\mathcal{M}_\kappa^d$. We set
\begin{equation}
\mathcal{F}_{\mu\nu}(\bbone):=F_{\mu\nu}
\end{equation}
with
\begin{equation}
F_{\mu\nu}=\mathcal{E}^{-2\gamma}\triangleright(X_\mu A_\nu-X_\nu A_\mu)+(\mathcal{E}^{1-\gamma}\triangleright A_\mu)\star(\mathcal{E}^{-\gamma}\triangleright A_\nu)-(\mathcal{E}^{1-\gamma}\triangleright A_\nu)\star(\mathcal{E}^{-\gamma}\triangleright A_\mu).\label{efmunu}
\end{equation}
It is easy to extend the (analog of) the field strength \eqref{efmunu} to a map
\begin{equation}
F:\mathbb{E}\to\mathbb{E}\otimes\Omega^2(\mathfrak{D}_\gamma)
\end{equation}
related to the $2$-form
\begin{equation}
F=\mathcal{E}^{-2\gamma}\triangleright {\bf{d}}A+\mathcal{E}^{-\gamma}\triangleright((\mathcal{E}\triangleright A)\times A)\label{curvat-f}.
\end{equation}
Finally, one can easily verify that $F$ \eqref{curvat-f} fulfills the Bianchi identity
\begin{equation}
{\bf{d}}F=(\mathcal{E}^{1+\gamma}\triangleright F)\times A-(\mathcal{E}^{2}\triangleright A)\times(\mathcal{E}^{\gamma}\triangleright F)\label{bianchi}.
\end{equation}

It is convenient to define the gauge group \cite{MW20} as 
\begin{equation}
\mathcal{U}:=\{g\in\mathbb{E},\ \ g^\dag\star g=g\star g^\dag=\bbone \}\label{unitar-group1},
\end{equation}
which characterizes the set of automorphisms of $\mathbb{E}$, say $\text{Aut}(\mathbb{E})$, preserving its structure of right-module and compatible with the canonical Hermitian structure on $\mathcal{M}_\kappa^d$. This latter is simply given by 
\begin{equation}
h(m_1,m_2)=m_1^\dag \star m_2,\label{}  
\end{equation}
for any $m_1,m_2\in\mathbb{E}$. Indeed, for any $\varphi\in\text{Aut}(\mathbb{E})$, $\varphi(m\star a)=\varphi(m)\star a$ holds for any $m\in\mathbb{E}$. Hence, $\varphi(\bbone\star a)=\varphi(a)=\varphi(\bbone)\star a$ showing that the action of any $\varphi\in\text{Aut}(\mathbb{E})$ on the algebra is fully determined by its action on the unit $\varphi(\bbone)$. Then, set $g:=\varphi(\bbone)$ and consider the compatibility condition with the Hermitian structure{\footnote{ Recall that a Hermitian structure is a sesquilinear form $h:\mathbb{E}\otimes\mathbb{E}\to\mathcal{M}_\kappa^d$ satisfying $h(m_1,m_2)^\dag=h(m_2,m_1)$ and $h(m_1\star a_1,m_2\star a_2)=a_1^\dag\star h(m_1,m_2)\star a_2$ for any $m_1,m_2\in\mathbb{E}$ and any $a_1,a_2\in\mathcal{M}_\kappa^d$  .}} given by $h(\varphi(m_1\star a_1),\varphi(m_2\star a_2))=h(m_1\star a_1,m_2\star a_2)$ for $m_1=m_2=\bbone$. The result \eqref{unitar-group1} follows.\\

The twisted gauge transformations are defined by
\begin{equation}
\nabla_{X_{\mu}}^g(a)=(\mathcal{E}^{\gamma+1}\triangleright g^\dag)\star\nabla_{X_{\mu}}(g\star a),
\quad\forall\,g\in\mathcal{U},\,\forall\,a\in\mathcal{M}_\kappa^d.
\label{gaugebitwist1}
\end{equation}
Accordingly, the gauge transformation of the gauge potential $A_\mu$ is
\begin{equation}
A_\mu^g=(\mathcal{E}^{\gamma+1}\triangleright g^\dag)\star A_\mu\star(\mathcal{E}^{\gamma}\triangleright g)+(\mathcal{E}^{\gamma+1}\triangleright g^\dag)\star X_\mu(g),\quad\forall\,g\in\mathcal{U},
\label{amugtwist}
\end{equation}
with $\nabla_{X_{\mu}}^g(a)=A_\mu^g\star(\mathcal{E}^{\gamma}\triangleright a)+X_{\mu}(a)$. The gauge transformation for the $1$-form connection reads
\begin{equation}
A^g=(\mathcal{E}^{\gamma+1}\triangleright g^\dag)\times A \star (\mathcal{E}^{\gamma}\triangleright g)+(\mathcal{E}^{\gamma+1}\triangleright g^\dag)\times {\bf{d}}g\label{gauge-trans-conn}.
\end{equation}
The corresponding gauge transformation for the field strength and $2$-form curvature are given for any $g\in\mathcal{U}$ by
\begin{equation}
F_{\mu\nu}^g=\mathcal{E}^2(g^\dag)\star F_{\mu\nu}\star g\quad\mbox{and}\quad F^g=\mathcal{E}^2(g^\dag)\times F\times g.\label{lebel1914}
\end{equation}
\\
From now on, we will focus on the case $\gamma=0$. The extension to nonzero values of $\gamma$ is straightforward and will not alter the conclusions obtained in the ensuing analysis.\\

\noindent At this point, one comment is in order:

It can be realized that the connection $\nabla_{X_{\mu}}$ does not satisfy the usual relation for Hermitian connection given by $h(\nabla_{X_{\mu}}(m_1),m_2)+h(m_1,\nabla_{X_{\mu}}(m_2))=X_\mu h(m_1,m_2)$, giving rise to $A_\mu=A^\dag_\mu$. This stems from the fact that the derivations are twisted derivations, in view of \eqref{tausigleibnitz}, and not real derivations since one has $(X_\mu(a))^\dag=\mathcal{E}^{-1}X_\mu(a^\dag)$. Instead, the relation is twisted as it can be expected. Indeed, a standard calculation yields 
\begin{equation}
h(\mathcal{E}^{-1}\triangleright\nabla_{X_{\mu}}(m_1),m_2)+h(\mathcal{E}^{-1}\triangleright m_1,\nabla_{X_{\mu}}(m_2))=X_\mu h(m_1,m_2)\label{cledag}
\end{equation}
for any $X_\mu\in\mathfrak{D}_\gamma$, $m_1,m_2\in\mathbb{E}$, provided $A_\mu=\mathcal{E}\triangleright A_\mu^\dag$. Note that a relation somewhat similar to \eqref{cledag} appears within nocommutative differential calculus based on $\varepsilon$-derivations and corresponding $\varepsilon$-connections \cite{JNG-eps}.

\subsection{The BRST symmetry.}

\label{section23}

We will follow the algebraic route used a long time ago in the context of the BRST symmetry for (commutative) Yang-Mills theories which gave rise to the algebraic theory of anomalies. Here, our purpose is to define algebraically the structure equations characterizing the BRST symmetry linked to the gauge symmetry \eqref{gauge-trans-conn} for $\gamma=0$ given by
\begin{equation}
A^g=(\mathcal{E}\triangleright g^\dag)\times A \star g+(\mathcal{E}\triangleright g^\dag)\times {\bf{d}}g,\ \ \ \label{gauge-trans-zero}
\end{equation}
together with \eqref{lebel1914}.\\

First, one easily defines from \eqref{gauge-trans-zero} the following operation
\begin{equation}
\delta_\omega A_\mu=X_\mu(\omega)+A_\mu\star \omega-(\mathcal{E}\triangleright \omega)\star  A_\mu,\ \ \delta_\omega A={\bf{d}}\omega+A\times \omega-(\mathcal{E}\triangleright \omega)\times  A\label{deltaomega}
\end{equation}
where $\omega \in\Omega^0(\mathfrak{D}_0)$, which can be viewed as the infinitesimal transformations linked to \eqref{gauge-trans-zero}. Recall that $A\in\Omega^1(\mathfrak{D}_0)$ denotes the $1$-form connection. One can verify that 
\begin{equation}
[\delta_{\omega_1},\delta_{\omega_2}]:=\delta_{\omega_1},\delta_{\omega_2}-\delta_{\omega_2},\delta_{\omega_1}=\delta_{[\omega_1,\omega_2]}\label{inter1}
\end{equation}
where $[\omega_1,\omega_2]=\omega_1\star\omega_2 - \omega_2\star\omega_1$, hence entailing the set of $\delta_\omega$ transformations with a structure of Lie algebra.

Furthermore, one can verify that the classical action
\begin{equation}
S_{\mathrm{cl}}=\int (F\times F^\dag):=\int d^5x\  F_{\mu\nu}\star  F_{\mu\nu}^\dag\label{class-action}
\end{equation}
is invariant under the gauge transformations \eqref{lebel1914} as well as under the operation \eqref{deltaomega}, namely
\begin{equation}
\delta_\omega S_{\mathrm{cl}}=0.\label{brool}
\end{equation}
Note by the way that gauge invariance which now translates into \eqref{brool} holds only whenever the $\kappa$-Minkowski space is $5$-dimensional, i.e. $d=5$, as a consequence of the analysis of \cite{MW20}.

Promoting $\omega$ to a Grassman variable, i.e. introducing the ghost field $C$, gives rise as usual to the BRST counterpart of \eqref{deltaomega} given by the following structure equations
\begin{eqnarray}
s_0A&=& -{\bf{d}}C-A\times C-(\mathcal{E}\triangleright C)\times A,\label{brsA}\\
s_0C&=&-C\times C\label{brsC},
\end{eqnarray}
which, combined with \eqref{curvat-f} with $\gamma=0$, giving
\begin{equation}
 F= {\bf{d}}A+(\mathcal{E}\triangleright A)\times A,\label{fgamma0}
\end{equation}
yields
\begin{equation}
s_0F=F\times C-(\mathcal{E}^2\triangleright C)\times F\label{transfoF},
\end{equation}
while the invariance of $S_{\mathrm{cl}}$ \eqref{brool} translates as expected into
\begin{equation}
s_0S_{\mathrm{cl}}=0\label{invar-brsclass}.
\end{equation}
Here, the operation $s$, sometimes called the Slavnov operation, satisfies
\begin{equation}
s_0^2=0,\label{raymond}
\end{equation}
as it can be easily verified by a simple calculation.

Notice that the BRST transformations given in \eqref{brsA} and \eqref{brsC} are {\it{formally}} similar to the BRST transformations for a commutative non-Abelian Yang-Mills theory, up to a twist operating in the third term of \eqref{brsA}. A similar comment applies to the BRST transformation for $F$ with a twist occuring in the second term of \eqref{transfoF}. This reflects merely the fact that the present twisted BRST symmetry \eqref{brsA} and \eqref{brsC} is related to a twisted gauge transformation corresponding to \eqref{deltaomega}. Note that $s_0$ acts on products of forms as
\begin{equation}
s_0(\rho\times\eta)=s(\rho)\times \eta+(-1)^{\delta(\rho)}\rho\times\eta\label{pasdetwist}.
\end{equation}

The above nilpotent operation $s_0$ gives rise to the functional Slavnov identity which permits one to control the UV behavior of the BRST invariant action $S_{\mathrm{cl}}$ (see \eqref{invar-brsclass}) after a suitable gauge-fixing obtained by adding a BRST-exact term. Namely, a suitable gauge-fixed action is
\begin{eqnarray}
S&=&S_{\mathrm{cl}}+s_0\int d^5x(\overline{C}^\dag\star\mathcal{E}^{-4}(X_\mu A_\mu))\\
&=&S_{\mathrm{cl}}+\int d^5x (b.(X_\mu A_\mu)-\overline{C}. X_\mu( sA_\mu)    )\label{fullaction}
\end{eqnarray}
where $X_\mu$ is given by \eqref{tausig-famil} with $\gamma=0$, the symbol ``$.$'' denotes the usual commutative product and we used the identity \cite{PW2018,PW2018bis}
\begin{equation}
\int d^5x\ (f\star g^\dag)(x)=\int d^5x\ f(x).\overline{g}(x)
\end{equation}
($\overline{g}$ is the complex conjugate of $g$)which holds for any Schwartz functions. Here, the real-valued fields $\overline{C}$ and $b$ are respectively the antighost field and the St\"uckelberg auxiliary field whose functional integration serves to implement the gauge-fixing condition. The respective ghost numbers are $-1$ and $0$. The BRST transformations of these fields are
\begin{equation}
s_0\overline{C}=b,\ \ s_0b=0.
\end{equation}
The action of $s_0$ on the fields gives rise to the functional Slavnov identity. Namely, by supplementing $S$ with source terms one introduces
\begin{equation}
\Gamma^0:=S+S_{\mathrm{source}},\ \ S_{\mathrm{source}}=\int d^5x\ (j_\mu\star s_0A_\mu+j\star s_0C).
\end{equation}
Then, one easily infers that the functionnal Slavnov identity is given by
\begin{equation}
\mathcal{S}\Gamma^0:=\int d^5x\ \big(\frac{\delta\Gamma^0}{\delta j_\mu(x)}\frac{\delta\Gamma^0}{\delta A_\mu(x)} + \frac{\delta\Gamma^0}{\delta j(x)}\frac{\delta\Gamma^0}{\delta C(x)}+b(x)\frac{\delta\Gamma^0}{\delta \overline{C}(x)}\big)=0\label{slavnovidentity}
\end{equation}
capturing the BRST symmetry of the theory at the tree level. This Slavnov identity should serve to control the UV behavior of the theory as well as the gauge invariance at each order in the perturbative expansion, $\Gamma^0$ being replaced by its renormalized counterpart at the given order. This will not be of our concern here.\\

\section{Twisted BRST symmetry and its related algebras.}

\label{section3}

It turns out that \eqref{brsA} and \eqref{brsC} can be related, in particularly relevant way to be described in a while, to a twisted BRST symmetry obtained from a noncommutative analog of a horizontality condition. Indeed, the nilpotent operation linked to this latter BRST symmetry acts as a twisted derivation exactly as ${\bf{d}}$ acts, justifying the terminology ``twisted BRST'',  while $s_0$ given in \eqref{brsA} and \eqref{brsC}, which characterizes an invariance of $S_{\mathrm{cl}}$ \eqref{class-action}, is a non-twisted derivation.

At this stage of the analysis, it is instructive to recall the main features of the algebraic framework ruling the BRST symmetry for commutative Yang-Mills theories, introducing in particular the notion of BRST algebra as a bigraded differential algebra encompassing the $1$-form connection, ghost field, related curvature and the BRST operation as one component of the total differential. This will be done in subsection \ref{section31}. This framework based on the notion of Weil algebra, whose basic features are summarized in the appendix \ref{weil}, is sufficiently universal and flexible to serve as a guideline to define a suitable characterization of the algebraic set-up  for the twisted BRST symmetry and its non-twisted partner.

Recall that the usefulness of the BRST symmetry for Yang-Mills theory goes much beyond the gauge-fixing procedure of the classical action functional. For a geometrical interpretation as well as related algebraic viewpoints{\footnote{The relevance of the Weil algebra in the BRST framework was initially suggested by R. Stora. }}, see \cite{Stora84}-\cite{dbv-weil}. Recall that the BRST symmetry plays a central role in the algebraic theory of perturbative anomalies for which ``solving'' the related Wess-Zumino Consistency Condition actually reduces to solve the $s$-cohomology modulo $\bf{d}$ \cite{Stora84}-\cite{dbv-tv}, while the higher order cocycles occurring in the corresponding descent equations are linked with a tower of anomalous correlation functions of the BRST current algebra \cite{wal87}. In the same way, the BRST symmetry is essential in Topological Field Theories \cite{rakow} of cohomological class to perform a suitable gauge fixing \cite{wal89} as well as in the determination of the corresponding invariants. This applies to the Donaldson invariants \cite{donald90} stemming from the $4$-dimensional topological Yang-Mills theory \cite{Wit88} as well as the invariants related to the $2$-dimensional topological gravity \cite{Wit88bis}, where in each case the use of a suitable BRST symmetry was shown \cite{STW94} to be essential to characterize the relations between the different schemes describing the equivariant cohomology \cite{GHV73} relevant to these theories.

\subsection{BRST algebra and Weil algebra in Yang-Mills theory.}

\label{section31}

In this subsection, we use mostly the notations of \cite{STW94}. Given a Lie algebra $\mathfrak{g}$, it is known that the structure of the BRST algebra for commutative Yang-Mills theories follows closely the generic structure of the Weil algebra \cite{cartan50,dbv-weil,STW94}. As we now recall, this is a mere modification of the material presented in the appendix \ref{weil}.\\

The relevant differential algebra is built from two copies of the algebra $\mathcal{W}(\mathfrak{g})$ \eqref{doubleV}. One copy is still $\mathcal{W}(\mathfrak{g})$, the free algebra generated by the $\omega^a$'s with degree $1$ and by the $\Omega^a$'s with degree 2. The other one, denoted by $\mathcal{W}_{\phi\pi}(\mathfrak{g})$ is the free algebra generated by additional elements $C^a$'s with new degree $1$ and by the $\phi^a$'s with new degree 2. $\{C^a\}_{a\in \mathcal{I}}$ and $\{\phi^a\}_{a\in \mathcal{I}}$ are therefore respectively basis of two additional copies of $\mathfrak{g}^*$, the dual of $\mathfrak{g}$ with basis $\{T^a\}_{a\in \mathcal{I}}$. As for \eqref{doubleV}, one has $\mathcal{W}_{\phi\pi}(\mathfrak{g})=\bigwedge\mathfrak{g}^*(C)\otimes\bigvee\mathfrak{g}^*(\phi)$ with $\bigwedge\mathfrak{g}^*(C)=\bigoplus_{g\in\mathbb{N}}\bigwedge^g\mathfrak{g}^*(C)$ (resp.$\bigvee\mathfrak{g}^*(\phi)=\bigoplus_{g\in\mathbb{N}}\bigvee^g\mathfrak{g}^*(\phi)$) where $\bigwedge^g\mathfrak{g}^*(C)$ (resp. $\bigvee^g\mathfrak{g}^*(\phi)$) involves forms of new degree $g$ (resp. $2g$), identified with the ghost number.

The elements $\omega^a$ and $\Omega^a$ still verify
 \eqref{abstractF}, \eqref{abstractbianchi} while $C^a$ and $\phi^a$ obey similar relations
\begin{equation}
\phi=sC+\frac{1}{2}[C,C]\label{abstractphi},
\end{equation}
with the Bianchi identity 
\begin{equation}
s\phi+[C,\phi]=0, \label{abstractCCCP}
\end{equation}
where $C=T^a\otimes C^a$, $\phi=T^a\otimes\phi^a$ and one has $s^2=0$. The differential $s$ will be identified with the Slavnov operation for the BRST symmetry.

From the two graded differential algebra{\footnote{We drop from now on the subscript $\mathcal{W}$ in ${\bf{d}}_{\mathcal{W}}$}} $(\mathcal{W}(\mathfrak{g}), {\bf{d}})$ and $(\mathcal{W}_{\phi\pi}(\mathfrak{g}), s)$, the BRST algebra can then be defined as the following differential algebra
\begin{equation}
\mathcal{W}_{\mathrm{BRST}}(\mathfrak{g})=\big(\mathcal{W}(\mathfrak{g})\otimes\mathcal{W}_{\phi\pi}(\mathfrak{g}), \widetilde{\bf{d}}={\bf{d}}+s,\widetilde{\omega}=A+C,\widetilde{\Omega}=\widetilde{\bf{d}}\widetilde{\omega}+\frac{1}{2}[\widetilde{\omega},\widetilde{\omega}]\big)\label{alg-brst},
\end{equation}
(further supplemented by the condition $\widetilde{\Omega}=\Omega$, see \eqref{horizont-yangmills} below), which is now bigraded, each element carrying a bidegree $(p,g)$ where $p$ (resp. $g$) is the degree of form (resp. ghost number){\footnote{The bigrading can be straighforwardly extended to a $\mathbb{Z}$-bigrading for convenience.}}. ${\bf{d}}$ and $s$ in the total differential $\widetilde{\bf{d}}$
are now sub-differentials with respective bidegree $(1,0)$ and another one with bidegree $(0,1)$. Using ${\bf{d}}^2=0$, $ s^2=0$ and 
\begin{equation}
s{\bf{d}}+{\bf{d}}s=0,
\end{equation}
one obtains $\widetilde{\bf{d}}^2=0$. The generators $\widetilde{\omega}=A+C$ and $\widetilde{\Omega}$ of $\mathcal{W}_{\mathrm{BRST}}(\mathfrak{g})$ satisfy relations similar to \eqref{abstractF}, \eqref{abstractbianchi} given by
\begin{equation}
\widetilde{\Omega}=\widetilde{\bf{d}}\widetilde{\omega}+\frac{1}{2}[\widetilde{\omega},\widetilde{\omega}],\ \ \widetilde{\bf{d}}\widetilde{\Omega}+[\widetilde{\omega},\widetilde{\Omega}]=0\label{tildeomegaSS}
\end{equation}
where the commutators are graded commutators w.r.t. the total degree defined by the sum of the form degree and the ghost number (mod $2$). The structure equations for the BRST symmetry for $\mathfrak{g}$ stem from the condition, sometimes called the ``Russian formula'' \cite{Stora84}:
\begin {equation}
\widetilde{\Omega}={\Omega}\label{horizont-yangmills},
\end{equation}
which, after expanding the LHS of \eqref{horizont-yangmills} in ghost numbers and identifying the terms of same ghost number in both sides yields 
\begin{equation}
s\omega=-{\bf{d}}C-[\omega,C],\ \ sC=-CC\label{yang-brst}.
\end{equation}

Algebraic connections on $\mathcal{W}_{\mathrm{BRST}}(\mathfrak{g})$ \eqref{alg-brst} are splitted into components of bidegree $(1,0)$ and $(0,1)$. If  $\widetilde{\omega}=A+C$ is such a connection, one then has for any $\lambda\in\mathfrak{g}$
\begin{equation}
i(\lambda)\widetilde{\omega}=\lambda,\ \ L(\lambda)\widetilde{\omega}=[\widetilde{\omega}, \lambda],
\end{equation}
in which, setting in obvious notation $ L_{\phi\pi}(\lambda)=i_{\phi\pi}(\lambda)s+s i_{\phi\pi}(\lambda)$, 
\begin{eqnarray}
i= i_{\mathcal{W}}+ i_{\phi\pi},\ \ 
L=i\widetilde{d}+\widetilde{d}i =L_{\mathcal{W}}+ L_{\phi\pi}\label{Ltotal},
\end{eqnarray}
stemming from 
\begin{equation}
 i_{\phi\pi}d+d i_{\phi\pi}=0,\ \  i_{\mathcal{W}}s+s i_{\mathcal{W}}=0\label{anticomut},
\end{equation}
where $ i_{\mathcal{W}}$ and $ i_{\phi\pi}$ carry respective bidegree $(-1,0)$ and $(0,-1)$ while the $L$'s have bidegree $(0,0)$. From the expansion in ghost numbers (bidegrees) of \eqref{Ltotal}, one easily obtains the action of $ i_{\mathcal{W}}$, $ L_{\mathcal{W}}$, $ i_{\phi\pi}$, $ L_{\phi\pi}$ on the various generators. Besides, one can check that $\widetilde{\Omega}$ fulfills
\begin{equation}
i(\lambda)\widetilde{\Omega}=0,\ \ L(\lambda)\widetilde{\Omega}=[\widetilde{\Omega},\lambda],\ \ \forall\,\lambda\in\mathfrak{g}\label{coubure-tilde}.
\end{equation}
Hence, $\widetilde{\Omega}$ occuring in \eqref{horizont-yangmills} is horizontal.

Note that, merely extending the notion of Weil algebra recalled in the appendix \ref{weil}, one can define the Weil algebra of the BRST symmetry for Yang-mill theories as $(\mathcal{W}_{\mathrm{BRST}}(\mathfrak{g}),i)$ where $i$ defined in the first relation \eqref{Ltotal}. 

\subsection{Twisted BRST symmetry from a Russian formula.}

\label{section32}

We now turn to the case of the BRST symmetry derived in subsection \ref{section23}. We will follow rather closely the various steps used in the appendix \ref{weil} and in subsection \ref{section31}, adapting when necessary the initial framework to the relevant noncommutative setting.\\

We start from the $1$-form connection $A\in\Omega^1(\mathfrak{D}_0)$ introduced in \eqref{cestamiou} and \eqref{cestamioubis} and its $2$-form curvature $F\in\Omega^2(\mathfrak{D}_0)$ defined in \eqref{curvat-f}. Define in a way somehow similar to \eqref{doubleV} the free graded algebra $\mathbb{W}(A)$ generated by $A$ and $F$ (respectively with degree $1$ and $2$). The use of degree of forms introduced in subsection \ref{section21} gives rise to $\mathbb{W}(A)=\bigoplus_{p\in\mathbb{N}}\mathbb{W}^p(A)$ where $p$ is the degree of form ($\mathbb{W}(A)\subset\Omega^\bullet(\mathfrak{D}_0)$). According to \eqref {curvat-f} and \eqref{bianchi}, $A$ and $F$ verify the by-now obvious relations
\begin{eqnarray}
F&=&{\bf{d}}A+(\mathcal{E}\triangleright A)\times A\label{gener1},\\
{\bf{d}}F&=&(\mathcal{E}\triangleright F)\times A-(\mathcal{E}^{2}\triangleright A)\times F\label{gener2}
\end{eqnarray}
which may be viewed as the analogs of \eqref{abstractF} and \eqref{abstractbianchi}. Equipping $\mathbb{W}(A)$ with the differential ${\bf{d}}$, which obeys ${\bf{d}}^2=0$, as signaled by the Bianchi identity \eqref{gener2}, turns $(\mathbb{W}(A),{\bf{d}})$ into a graded differential algebra which however is no longer commutative, a change which will play no role in the ensuing analysis. Recall that ${\bf{d}}$ is a twisted differential with twists $(\bbone,\mathcal{E})$ in view of \eqref{tausigleibnitz}, i.e. \eqref{leibnitz-form} yields for $\gamma=0$
\begin{equation}
{\bf{d}}(\rho\times\eta)={\bf{d}}\rho\times\eta+(-1)^{\delta(\rho)}(\mathcal{E}\triangleright\rho)\times{\bf{d}}\eta,
\quad\forall\,\rho,\eta\in\mathbb{W}\left(A\right)
\label{leibnitz-form-ZERO},
\end{equation}
which extends to $\rho,\eta\in\Omega(\mathfrak{D}_0)$, where $\delta(.)$ is still the form degree as in subsection \ref{section21}. Now, define one copy of $(\mathbb{W}(A),{\bf{d}})$, hereafter denoted by $(\mathbb{W}(C),s_1)$, by introducing new generators $C$ and $\phi$ playing respectively the role of $A$ and $F$ and a new twisted differential $s_1$ with twists $(\bbone,\mathcal{E})$, that is, $s_{1}$ satisfies 
\begin{equation}
s_1(\rho\times\eta) = s_1\rho\times\eta+(-1)^{\delta(\rho)}(\mathcal{E}\triangleright\rho)\times s_1\eta,
\quad\forall\,\rho,\eta\in\mathbb{W}\left(C\right)
\label{leibnitz-form-ZERO}
\end{equation}
(which extends to $\rho,\eta\in\Omega^{\bullet\prime}(\mathfrak{D}_0)$, one copy of $\Omega^\bullet(\mathfrak{D}_0))$.

The new degree resulting from this construction is identified with the ghost number, as in subsection \ref{section31}. One has $\mathbb{W}(C)=\bigoplus_{g\in\mathbb{N}}\mathbb{W}^g(C)$ with $g$ being the ghost number. The generators $C$ and $F$ verify by construction
\begin{eqnarray}
\phi&=&s_1C+(\mathcal{E}\triangleright C)\times C\label{gener3},\\
s_1\phi&=&(\mathcal{E}\triangleright \phi)\times C-(\mathcal{E}^{2}\triangleright C)\times \phi\label{gener4},
\end{eqnarray}
and the twisted differential $s_1$ with twists $(\bbone,\mathcal{E})$ is such that $s_1^2=0$ in view of \eqref{gener4}. $(\mathbb{W}(C),s_1)$ is again a graded differential algebra and it can be easily seen that the data  $(\widehat{\mathbb{W}}=\mathbb{W}(A)\otimes\mathbb{W}(C),\widehat{{\bf{d}}}_1)$ is a bigraded differential algebra where the total twisted differential $\widehat{{\bf{d}}}_1={\bf{d}}+s_1$ with twists $(\bbone,\mathcal{E})$. ${\bf{d}}$ and $s_1$ are differentials of respective bidegree $(1,0)$ and $(0,1)$. To see that, one first notices that forms in $\widehat{\mathbb{W}}$ have a bidegree $(p,g)$ where $p$ (resp. $g$) is the form degree (resp. ghost number).  For instance, $A$ carries a bidegree $(1,0)$ while $C$ has bidegree $(0,1)$. Since $\mathbb{W}(A)$ and $\mathbb{W}(C)$ are each a subalgebra of $\Omega^\bullet(\mathfrak{D}_0)$, it is convenient to view $\widehat{\mathbb{W}}$ as a subalgebra of the bigraded algebra $\widehat{\Omega}$ built from these two copies of $\Omega^\bullet(\mathfrak{D}_0)$. One then can write $\widehat{\mathbb{W}}\subset\widehat{\Omega}=\bigoplus_{p,g}\widehat{\Omega}^{p,g}(\mathfrak{D}_0)$ and the differentials ${\bf{d}}$ and $s_1$ extend to maps such that 
\begin{eqnarray}
{\bf{d}}&:&\Omega^{p,g}(\mathfrak{D}_0)\to\Omega^{p+1,g}(\mathfrak{D}_0),\label{derviD}\\
s_1&:&\Omega^{p,g}(\mathfrak{D}_0)\to\Omega^{p,g+1}(\mathfrak{D}_0)\label{derivs}.
\end{eqnarray}
Note that the product of forms verifies $\widehat{\Omega}^{p_1,g_1}(\mathfrak{D}_0)\times\widehat{\Omega}^{p_2,g_2}(\mathfrak{D}_0)\subset\widehat{\Omega}^{p_1+p_2,g_1+g_2}(\mathfrak{D}_0)$.\\
Next, for any $\alpha\in\widehat\Omega$ with form degree $p$ and ghost number $g$, define the total degree $|\alpha|$ as
\begin{equation}
|\alpha|:= \delta(\alpha)+g\label{totgrad}.
\end{equation}
in which $\delta(\alpha)$ is the form-degree of $\alpha$, already introduced below eqn. \eqref{leibnitz-form}. The Leibniz rule obeyed by ${\bf{d}}$ and $s_1$ then extends for any $\rho,\eta\in\widehat{\Omega}$ as
\begin{eqnarray}
s_1(\rho\times\eta)&=&s_1(\rho)\times\eta+(-1)^{|\rho|}(\mathcal{E}\triangleright\rho)\times s_1(\eta),\label{leibs1}\\
{\bf{d}}(\rho\times\eta)&=&{\bf{d}}\rho\times\eta+(-1)^{|\rho|}(\mathcal{E}\triangleright\rho)\times{\bf{d}}\eta\label{leibdd}.
\end{eqnarray}
Now, define
\begin{equation}
{\widehat{{\bf{d}}}}_1={\bf{d}}+s_1\label{totaldifferencial}
\end{equation}
which satisfies
\begin{equation}
{\widehat{{\bf{d}}}}_1^2=0\label{carretotdiff}
\end{equation}
whenever one has
\begin{equation}
s_1{\bf{d}}+{\bf{d}}s_1=0\label{sdanticomm},
\end{equation}
which holds here. It follows that ${\widehat{{\bf{d}}}}_1$ can be interpreted as the total differential which can equip  $\widehat{\mathbb{W}}\subset\widehat{\Omega}$. Processing in analogy with subsection \ref{section31}, we now introduce
\begin{eqnarray}
 \widehat{A}&=&A+C\label{tildavar},\\
\widehat{F}&=&{\widehat{{\bf{d}}}}_1\widehat{A}+\frac{1}{2}\langle \widehat{A},\widehat{A} \rangle,\label{tildeF}
\end{eqnarray}
where the graded twisted commutator is given by
\begin{equation}
\langle \rho,\eta\rangle:=(\mathcal{E}^{|\rho|\delta(\eta)}\triangleright\rho)\times \eta-(-1)^{|\rho||\eta|}(\mathcal{E}^{|\eta|\delta(\rho)}\triangleright\eta)\times\rho,\ \ \label{twist-commut}
\end{equation}
for any bigraded forms $\rho,\eta$ with $|\rho|=|\eta|=1$. \\

Then, the data $(\widehat{\mathbb{W}},{\widehat{{\bf{d}}}}_1, \widehat{A},\widehat{F})$ with $\widehat{F}$ verifying
\begin{equation}
\widehat{F}=F\label{YpaYpa},
\end{equation}
can be viewed as a noncommutative analog of the BRST algebra. The expansion of the LHS of  the ``Russian formula'' \eqref{YpaYpa} in ghost numbers, using
\begin{equation}
\langle A,A\rangle=2(\mathcal{E}\triangleright A)\times A,\ \ \langle A,C\rangle=\langle C,A\rangle=A\times C+(\mathcal{E}\triangleright C)\times A,\ \ \langle C, C\rangle =2C\times C\label{lescommuto},
\end{equation}
yields the following structure equations
\begin{align}
    s_{1}C &= -(\mathcal{E}\triangleright C) \times C\label{s1C},\\
    s_{1}A &= -{\bf{d}}C - (\mathcal{E}\triangleright C)\times A - (\mathcal{E}\triangleright A)\times C\label{s1A}.
\end{align}
The curvature $\widehat{F}$ of the algebraic connection $\widehat{A}$ satisfies the Bianchi identity
\begin{align}
    \widehat{d}_1\widehat{F}
    = (\mathcal{E}\triangleright \widehat{F})\times \widehat{A}
    - (\mathcal{E}^{2}\triangleright \widehat{A})\times \widehat{F}\label{decadix-develour}
\end{align}
which combined with \eqref{YpaYpa} gives rise to
\begin{equation}
s_{1}F =(\mathcal{E}\triangleright F)\times C - (\mathcal{E}^{2}\triangleright C)\times F.\label{triphase}
\end{equation}
Summarizing the above analysis, it is natural to define the noncommutative analog of the BRST algebra for Yang-Mills theory \eqref{alg-brst} by the data
\begin{equation}
(\widehat{\mathbb{W}},{\widehat{{\bf{d}}}}_1={\bf{d}}+s_1, \widehat{A}=A+C,\widehat{F}={\widehat{{\bf{d}}}}_1\widehat{A}+\frac{1}{2}\langle \widehat{A},\widehat{A} \rangle),\label{NCbrst-alg}
\end{equation}
with $\widehat{F}\in\Omega^{2,0}(\mathfrak{D}_0)$. Since $s_1$ acts as a twisted derivation \eqref{leibs1}, this differential algebra actually corresponds to a twisted BRST symmetry which we will call the twisted BRST symmetry algebra. \\

However, assuming now that $d=5$ (i.e. working with $\mathcal{M}_\kappa^5$), it appears that the nilpotent operation $s_1$ is not a symmetry of the classical action $S_{\mathrm{cl}}$ \eqref{class-action}, i.e. $s_1S_{\mathrm{cl}}\ne0$, while $s_0S_{\mathrm{cl}}=0$. This can be easily verified by combining \eqref{class-action} with \eqref{triphase}. It would be tempting to simply replace into the above construction $s_1$ by $s_0$ defined in \eqref{brsA} and \eqref{brsC}, but this would change the RHS of ${\widehat{{\bf{d}}}}_1$ \eqref{totaldifferencial} into ${{{\bf{d}}}}+s_0$ where ${{{\bf{d}}}}$ has twists $(\bbone,\mathcal{E})$ while $s_0$ is non-twisted in view of \eqref{pasdetwist}, implying that ${{{\bf{d}}}}+s_0$ is no longer a derivation. Therefore, ${{{\bf{d}}}}+s_0$ will never be a differential and the resulting $(\widehat{\mathbb{W}},{{{\bf{d}}}}+s_0)$ fails to be a differential algebra.\\

Nevertheless, it turns out that the BRST operation $s_0$ representing a symmetry of the classical action can be rigidly linked to the twisted BRST operation $s_1$. We now show that $s_1$ can be continously deformed into $s_0$, as elements of $\text{End}(\widehat\Omega)$. \\
First, notice that the use of a deformation of the form $\left\lbrace s_{t} = \left(1 - t\right)s_{0} + ts_{1},\,t\in\left[0,1\right]\right\rbrace$ is unsuitable for our purpose. Indeed, the elements of this $1$-parameter family do not satisfy a (possibly twisted) Leibniz rule.\\

Instead of this $1$-parameter family, consider the $1$-parameter family $\left\lbrace s_{t},\,t\in\left[0,1\right]\right\rbrace$ such that, for any $t\in\left[0,1\right]$, $s_{t}$ satisfies a twisted Leibniz rule with twists $\left(\bbone,\mathcal{E}^t\right)$, i.e.
\begin{align}
    s_{t}\left(\rho\star \eta\right) 
    &= s_{t}\left(\rho\right)\times \eta 
    + \left(-1\right)^{\left|\rho\right|}(\mathcal{E}^{t}\triangleright\rho)\star s_{t}\left(b\right),
    \quad\forall\,t\in[0,1],\,\forall\,\rho,\eta\in\widehat{\Omega}
    \label{leibst}
    \end{align}
The action of $s_t$ on the generators is defined by
\begin{align}
    s_{t}C &= -(\mathcal{E}^{t}\triangleright C)\times C,\label{stC}\\
    s_{t}A &= -{\bf{d}}C - (\mathcal{E}\triangleright C)\times A - (\mathcal{E}^{t}\triangleright A)\times C\label{stA},
\end{align}
which implies in particular that
\begin{equation}
	s_{t}F = (\mathcal{E}^{t}\triangleright F)\times C - (\mathcal{E}^{2}\triangleright C)\times F.
\end{equation}
One easily check that $s_{t=0}=s_0$ and $s_{t=1}=s_1$. Note that \eqref{stC} and \eqref{stA} are formally obtained from the expansion in ghost numbers of $\widehat{F}_t = F$ in which  $\widehat{F}_t$ is defined by
\begin{align}
    \widehat{F}_t
    = \widehat{\bf{d}}_{t}\widehat{A} + \frac{1}{2}\langle\widehat{A},\widehat{A}\rangle_{t}\label{horizont-bizar}
\end{align}
with $\widehat{\bf{d}}_{t} = {\bf{d}} + s_{t}$ and 
\begin{align}
 \langle A,A\rangle_t &= 2(\mathcal{E}\triangleright A)\times A,\ \ \langle C,C\rangle_{t} = 2(\mathcal{E}^{t}\triangleright C)\times C,\\
  \langle A,C\rangle_{t} &= \langle C,A\rangle_{t} 
    = (\mathcal{E}\triangleright C)\times A + (\mathcal{E}^{t}\triangleright A)\times C.
\end{align}
Furthermore, it can be verified by a standard computation that
\begin{equation}
	s_{t}^{2} = 0,\ \ s_t{\bf{d}}+{\bf{d}}s_t=0,\ \ \forall\,t\in[0,1].
\end{equation}
Notice that $\widehat{\bf{d}}_{t}$ still verifies $\widehat{\bf{d}}_{t}^2=0$ but is plainly not a derivation. Notice also that $s_t$ is not an invariance of the classical action, i.e. $s_{t}S_{\mathrm{cl}}\ne0$, unless $t=0$.\\

Let us comment the above analysis. It appears that \eqref{NCbrst-alg} can be viewed as the natural algebraic structure describing a noncommutative analog of the BRST algebra \eqref{alg-brst}. Furthermore, since $\bf{d}$ and $s_1$ carry the same twists $(\bbone,\mathcal{E})$, $\widehat{\bf{d}}_1$ is a differential which, as such, is nilpotent {\it{and}} obeys in particular a twisted Leibniz rule. This is why a Bianchi identity \eqref{decadix-develour} of usual form for the total curvature \eqref{tildeF} holds true (basically, it involves only hatted variables). This is no longer valid for $F_t$ \eqref{horizont-bizar} stemming from $\widehat{\bf{d}}_t={\bf{d}}+s_t$, for any $t\in[0,1[$ since $\widehat{\bf{d}}_t$, albeit still nilpotent, does not obey a twisted Leibniz rule, implying the occurrence of unwanted terms in the counterpart of \eqref{decadix-develour}. Note that since there is basically no standard Bianchi identity for $F_t$, $t\ne1$, this latter quantity cannot be viewed as a curvature so that the algebraic status of the condition $\widehat{F}_t = F$, $t\ne1$, giving rise to \eqref{stC}, \eqref{stA} is unclear. \\
Finally, remark that the deformation we exhibit here is unique by contruction. It could be regarded as a homotopy in $\mathrm{End}(\widehat{\Omega})$. However, this terminology is not quite appropriate, as it implies somehow forgetting at least the Leibniz rule, which is fundamental for the present work. If, for $t\in\left[0,1\right]$, we call $L_{t}$ the subspace of $\mathrm{End}(\widehat{\Omega})$, consisting in the elements that anticommute with $\bf{d}$ and satisfy the twisted Leibniz rule with the twists $\left(\bbone,\mathcal{E}^t\right)$, then our path $\left\lbrace s_{t},\,t\in\left[0,1\right]\right\rbrace$ actually crosses once and only once each of these $L_{t}$ transversally, i.e., at exactly one point (and for each $t\in\left[0,1\right]$, this point is the operator $s_{t}$, which is nilpotent), while we would require the deformation to happen in the same subspace to call it appropriately a homotopy.

\section{Conclusion.}

\label{section4}

To summarize, we have exhibited a (continuous) map defined by $s_t:[0,1]\times\widehat{\Omega}\to\widehat{\Omega}$, \eqref{leibs1}-\eqref{stA} and satisfying $s_{t=1}=s_1$ given by \eqref{s1C}, \eqref{s1A} together with $s_{t=0}=s_0$ given by \eqref{brsA}, \eqref{brsC}. This map defines a path in $\mathrm{End}(\widehat{\Omega})$ such that each point of this path is a twisted derivation which is nilpotent and anticommutes with ${\bf{d}}$. In other words, although the data $(\widehat{W},\widehat{\bf{d}}_{0}={\bf{d}}+s_0)$ cannot be promoted to a differential algebra ($\widehat{\bf{d}}_{0}$ is not a differential!) so that it cannot be reliably viewed as a BRST algebra, it can be deformed to a full differential algebra given by \eqref{NCbrst-alg}, preserving nilpotency and anticommutativity with ${\bf{d}}$ and twisting continuously the Leibniz rule, through the map $s_{t}$. In some sense, the gauge invariance of the classical action represented by $s_0$ is rigidely (by uniqueness of $s_{t}$) linked by this map to the twisted BRST symmetry algebra \eqref{NCbrst-alg}. This results in two nilpotent operations. One, $s_0$ presented in subsection \ref{section23}, is actually the noncommutative analog of the (historical) Slavnov operation directly related to field theory as generating the Slavnov-Taylor identities \eqref{slavnovidentity} controlling its perturbative behavior. The other one, $s_1$ defined in subsection \ref{section32}, is actually the ghost number $+1$ component of the total differential of the twisted BRST symmetry algebra and as such is rigidely linked to the corresponding algebraic structure, stemming in particular from a noncommutative analog of the Russian formula.

The immediate application of the present framework is the use of $s_0$ and its functional Slavnov-Taylor identity to investigate quantum properties of the $\kappa$-Poincar\'e invariant gauge theory constructed in \cite{MW20}. This will be considered in a forthcoming publication.

\vskip 1 true cm

\noindent{\bf{Acknowledgments:}} Ph. M. thanks Stephan Stolz for answering some key questions related to this work. J.-C. W. thanks the Action CA18108 QG-MM from the European Cooperation in Science and Technology (COST).

\appendix

\section{The Weil algebra of a Lie algebra.}

\label{weil}

The notion of Weil algebra is a flexible algebraic tool encompassing various situations whenever an action (of e.g. a Lie algebra) comes into play. This action can be translated into the action of suitable Cartan operations \cite{GHV73}. The simplest example is provided by the case of a principal fiber bundle with connection, with the algebra of differential forms on the fiber bundle as relevant differential algebra and usual Cartan operations for the Lie algebra of the structure group acting on forms (and in particular  the form connection) describing the Lie algebra action on the fiber bundle.\\

A convenient way to describe the Weil algebra of a Lie algebra goes as follows. For a complete mathematical presentation, see chap. 6 in \cite{GHV73}. In this appendix, we use mostly the notations of ref. \cite{STW94}. Let $G$ be a Lie group with Lie algebra $\mathfrak{g}$ and dual space $\mathfrak{g}^*$. Let $\{T^a\}_{a\in \mathcal{I}}$ be the basis of $\mathfrak{g}$ and let $\{\omega^a\}_{a\in \mathcal{I}}$ and $\{\Omega^a\}_{a\in \mathcal{I}}$ denote respectively the basis of $\mathfrak{g}^*(\omega)$ and $\mathfrak{g}^*(\Omega)$, two copies of $\mathfrak{g}^*$. Consider the free (graded commutative) algebra $\mathcal{W}(\mathfrak{g})$ generated by the $\omega^a$'s with degree $1$ and by the $\Omega^a$'s with degree 2 . Namely, one has 
\begin{equation}
\mathcal{W}(\mathfrak{g})=\bigwedge\mathfrak{g}^*(\omega)\otimes\bigvee\mathfrak{g}^*(\Omega)\label{doubleV}
\end{equation}
where $\bigwedge\mathfrak{g}^*(\omega)=\bigoplus_{p\in\mathbb{N}}\bigwedge^p\mathfrak{g}^*(\omega)$ denotes as usual the exterior algebra on $\mathfrak{g}$ and $\bigvee\mathfrak{g}^*(\Omega)=\bigoplus_{p\in\mathbb{N}}\bigvee^p\mathfrak{g}^*(\Omega)$ is the symmetric algebra on $\mathfrak{g}$. $\bigwedge^p\mathfrak{g}^*(\omega)$ involves forms of degree $p$ while $\bigvee^p\mathfrak{g}^*(\Omega)$ involves elements of degree $2p$, which therefore define the grading{\footnote{This $\mathbb{N}$-grading can be easily extended to a $\mathbb{Z}$-grading for further convenience (thus allowing in particular to include objects with negative ghost numbers). }}.\\
{\it{Define}} the following elements of $\mathfrak{g}\otimes\mathcal{W}(\mathfrak{g})$ $\omega:=T^a\otimes\omega^a,\ \ \Omega:=T^a\otimes\Omega^a$
which obey
\begin{equation}
\Omega={\bf{d}}_{\mathcal{W}}\omega+\frac{1}{2}[\omega,\omega]\label{abstractF}
\end{equation}
together with the Bianchi identity 
\begin{equation}
{\bf{d}}_{\mathcal{W}}\Omega+[\omega,\Omega]=0, \label{abstractbianchi}
\end{equation}
and ${\bf{d}}_{\mathcal{W}}^2=0$ which permits one to identify the derivation ${\bf{d}}_{\mathcal{W}}$ with degree $1$ as a differential on $\mathcal{W}(\mathfrak{g})$. In \eqref{abstractF}, \eqref{abstractbianchi} the commutator and differential satisfy for any $u^a,v^a\in\mathcal{W}(\mathfrak{g})$ $[T^a\otimes u^a,T^b\otimes v^b]=[T^a,T^b](u^a.v^b)$ and $\ {\bf{d}}_{\mathcal{W}}(T^a\otimes u^a)=T^a\otimes{\bf{d}}_{\mathcal{W}}(u^a)$ where the symbol $.$ in the first relation denotes the product of forms.  \\

To make contact with the Weil algebra, we now introduce the Cartan operations  \cite{cartan50}, \cite{GHV73} characterizing the action of $\mathfrak{g}$ on $(\mathcal{W}(\mathfrak{g}),{\bf{d}}_{\mathcal{W}})${\footnote{Indeed, there is a natural action of $\mathfrak{g}$ on itself through the adjoint representation, i.e. any $\lambda\in\mathfrak{g}$ acts on $\mathfrak{g}$ as $\text{Ad}_\lambda=[\lambda,.]$. This induces an action of $\mathfrak{g}$ on $\mathfrak{g}^*$ by duality of linear spaces. This action extends to the exterior algebra $\bigwedge\mathfrak{g}^*(\omega)$, hence on the above differential algebra. }}. These are $i_\mathcal{W}(\lambda)$ and $L_\mathcal{W}(\lambda)=i_\mathcal{W}(\lambda){\bf{d}}_{\mathcal{W}}+{\bf{d}}_{\mathcal{W}}i_\mathcal{W}(\lambda)$ for any $\lambda\in\mathfrak{g}$, respectively the inner product and the Lie derivative \cite{GHV73}. $i_\mathcal{W}(\lambda)$ acts as a derivation of degree $-1$. $i_\mathcal{W}(\lambda)$ maps $n$-forms into $(n-1)$ forms; in particular, for any $1$-form $\eta$ in $\mathcal{W}(\mathfrak{g})$, $i_\mathcal{W}(\lambda)\eta=\eta(\lambda)$. Then, one easily infers that $L_\mathcal{W}(\lambda) L_\mathcal{W}(\rho)-L_\mathcal{W}(\rho) L_\mathcal{W}(\lambda)=L_\mathcal{W}({[\lambda,\rho]})$ and $L_\mathcal{W}(\lambda)
 i_\mathcal{W}(\rho)-i_\mathcal{W}(\rho) L_\mathcal{W}(\lambda)=i_\mathcal{W}({[\lambda,\rho]})$.\\
Now, assume that the $1$-form $\omega$ is an algebraic connection \cite{GHV73}. As such, it verifies \cite{GHV73}
\begin{eqnarray}
i_\mathcal{W}(\lambda)\omega=\lambda,\ \ 
L_\mathcal{W}(\lambda)\omega=[\omega,\lambda]\label{isurA}.
\end{eqnarray}
for any $\lambda\in\mathfrak{g}$. Combining \eqref{isurA} with \eqref{abstractF}, one easily obtains
\begin{eqnarray}
i_\mathcal{W}(\lambda){\Omega}&=&0\label{horizontal},\\
L_\mathcal{W}(\lambda){\Omega}&=&[{\Omega},\lambda].\label{equivarF}
\end{eqnarray}
for any $\lambda\in\mathfrak{g}$. The first (resp. second) relation \eqref{isurA} is a mere generalization of the ``vertical condition'' (resp. equivariance condition) satisfied by a connection on a principal fiber bundle. The relation \eqref{horizontal} signals that the $2$-form $\Omega$ is horizontal. Indeed, pick a principal fiber bundle $P(M,G):=P$ over a manifold $M$ with structure group $G$ of Lie algebra $\mathrm{Lie}\left(G\right)$. Then a $\mathrm{Lie}\left(G\right)$-valued $1$-form connection $A$ on $P$ is such that $i_P(X)A=X$, $L_P(X)A=[A,X]$ for any $X\in\mathrm{Lie}\left(G\right)$ while its curvature $F$ is horizontal, namely $i_P(X)F=0$.

The Weil algebra of $\mathfrak{g}$ can be defined as $((\mathcal{W}(\mathfrak{g}),{\bf{d}}_{\mathcal{W}}),i_\mathcal{W})$. The differential algebra $(\mathcal{W}(\mathfrak{g}),{\bf{d}}_{\mathcal{W}})$ generalizes the algebra of forms on the fiber bundle $P$, while the Cartan operation $i_\mathcal{W}(\lambda)$ generalizes the action of $\mathrm{Lie}\left(G\right)$ on $P$ and on the related connection.

\section{$\kappa$-Poincar\'e algebra and deformed translations.}\label{apendixB}

We use the bicrossproduct basis \cite{majid-ruegg} in this paper. Let $\Delta:\mathcal{P}^d_\kappa\otimes\mathcal{P}^d_\kappa\to\mathcal{P}^d_\kappa$, $\epsilon:\mathcal{P}^d_\kappa\to\mathbb{C}$ and ${\bf{S}}:\mathcal{P}^d_\kappa\to\mathcal{P}^d_\kappa$, be respectively the coproduct, counit and antipode endowing $\mathcal{P}^d_\kappa$ with a Hopf algebra structure. Let $(P_i, N_i,M_i, \mathcal{E},\mathcal{E}^{-1})$, $i=1, 2, \hdots, d-1$, be respectively the momenta, boosts, rotations and $\mathcal{E}:=e^{-P_0/\kappa}$ generating the Lie algebra
\begin{equation}
[M_i,M_j]= i\epsilon_{ij}^{\hspace{5pt}k}M_k,\ [M_i,N_j]=i\epsilon_{ij}^{\hspace{5pt}k}N_k,\ [N_i,N_j]=-i\epsilon_{ij}^{\hspace{5pt}k}M_k\label{poinc1}, 
\end{equation}
\begin{equation}
[M_i,P_j]= i\epsilon_{ij}^{\hspace{5pt}k}P_k,\ [P_i,\mathcal{E}]=[M_i,\mathcal{E}]=0,\ [N_i,\mathcal{E}]=\frac{i}{\kappa}P_i\mathcal{E}\label{poinc2},
\end{equation}
\begin{equation}
[N_i,P_j]=-\frac{i}{2}\delta_{ij}\left(\kappa(1-\mathcal{E}^{2})+\frac{1}{\kappa}\vec{P}^2\right)+\frac{i}{\kappa}P_iP_j\label{poinc3}.
\end{equation}
The Hopf algebra structure is defined by
\begin{align}
\Delta P_0&=P_0\otimes\bbone+\bbone\otimes P_0,\ \Delta P_i=P_i\otimes\bbone+\mathcal{E}\otimes P_i,\ \Delta \mathcal{E}=\mathcal{E}\otimes\mathcal{E}\label{hopf1},\\
\Delta M_i&=M_i\otimes\bbone+\bbone\otimes M_i,\ \Delta N_i=N_i\otimes \bbone+\mathcal{E}\otimes N_i-\frac{1}{\kappa}\epsilon_{i}^{\hspace{2pt}jk}P_j\otimes M_k,\label{hopf2}\\
\epsilon(P_0)&=\epsilon(P_i)=\epsilon(M_i)=\epsilon(N_i)=0,\  \epsilon(\mathcal{E})=1\label{hopf3},\\
{\bf{S}}(P_0)&=-P_0,\ {\bf{S}}(\mathcal{E})=\mathcal{E}^{-1},\  {\bf{S}}(P_i)=-\mathcal{E}^{-1}P_i,\  {\bf{S}}(M_i)=-M_i,\\
{\bf{S}}(N_i)&=-\mathcal{E}^{-1}(N_i-\frac{1}{\kappa}\epsilon_{i}^{\hspace{2pt}jk}P_jM_k)\label{hopf4bis}.
\end{align}
Recall that the $\kappa$-Minkowski space $\mathcal{M}_\kappa^d$ can be viewed as the dual of the Hopf subalgebra $\mathcal{T}_\kappa^d$ generated by $P_\mu$, $\mathcal{E}$, the deformed translation algebra, which inherits a structure of $^*$-Hopf algebra through: $P_\mu^\dag=P_\mu$, $\mathcal{E}^\dag=\mathcal{E}$. Then, the following relation holds true
\begin{equation}
(t\triangleright f)^\dag={\bf{S}}(t)^\dag\triangleright f^\dag,\label{pairing-involution}
\end{equation}
for any $t$ in $\mathcal{T}_\kappa^d$ and any $f\in\mathcal{M}^d_\kappa$, implying
\begin{equation}
(P_0\triangleright f)^\dag=-P_0\triangleright(f^\dag),\ (P_i\triangleright f)^\dag=-\mathcal{E}^{-1}P_i\triangleright(f^\dag),\ (\mathcal{E}\triangleright f)^\dag=\mathcal{E}^{-1}\triangleright(f^\dag)\label{dag-hopfoperat}.
\end{equation}
The action of $\mathcal{T}_\kappa^d$ on $\mathcal{M}_\kappa^d$ is 
\begin{equation}
(\mathcal{E}\triangleright f)(x)=f(x_0+\frac{i}{\kappa},\vec{x}),\ \ 
(P_\mu\triangleright f)(x)=-i(\partial_\mu f)(x)\label{left-module1}.\\
\end{equation}

\end{document}